\documentclass[prd,longbibliography,nofootinbib,preprintnumbers,twocolumn,showpacs,floatfix,superscriptaddress]{revtex4-1}

\usepackage[utf8]{inputenc}      
\usepackage{amsmath}
\usepackage{amssymb}
\usepackage{graphicx}
\usepackage{color}
\usepackage{hyperref}

\newcommand{\bitem}{\begin{itemize}}
\newcommand{\eitem}{\end{itemize}}
\newcommand{\bwt}{\begin{widetext}}
\newcommand{\ewt}{\end{widetext}}
\newcommand{\beq}{\begin{equation}}
\newcommand{\eeq}{\end{equation}}

\newcommand{\bdm}{\begin{displaymath}}
\newcommand{\edm}{\end{displaymath}}
\newcommand{\bea}{\begin{eqnarray}}
\newcommand{\eea}{\end{eqnarray}}

\begin{document}
\jot = 1.4ex         

\title{Neutrinoless double beta decay: neutrino mass versus new physics}

\author{Gia Dvali$^{1,2}$, Alessio Maiezza$^{3}$, Goran Senjanovi\'c$^{1,4}$ and Vladimir Tello$^{4}$}\noaffiliation

\affiliation{
Arnold Sommerfeld Center, Ludwig-Maximilians University,  Munich, Germany\\
$^2$
Max-Planck Institute for Physics, Munich, Germany\\
$3$
Dipartimento di Scienze Fisiche e Chimiche, Universit\`a, degli Studi dell'Aquila, via Vetoio, I-67100 L'Aquila, Italy\\
$^4$
International Centre for Theoretical Physics, Trieste, Italy
}

\begin{abstract}
Neutrinoless double beta decay is the textbook example of lepton number violation, often claimed to be a probe of neutrino Majorana mass. However, it could be triggered by new physics; after all, neutrino Majorana mass requires physics beyond the Standard Model. If at least one electron were right-handed, it would automatically signify new physics rather than neutrino mass. In case both electrons were left-handed, the situation would become rather complicated, and additional effort would be needed to untangle the source for this process. We offer a comprehensive study of this issue from both the effective operator approach and the possible UV completions, including the Pati-Salam quark-lepton unification. While neutrino exchange is natural and physically preferred, our findings show that new physics can still be responsible for the neutrinoless double beta decay. In particular, the Pati-Salam theory can do the job, consistently with all the phenomenological and unification constraints, as long as the unification scale lies above $10^{12} \text{GeV}$, albeit at the price of fine-tuning of some scalar masses.
\end{abstract}


\maketitle

\section{Introduction} \label{Introduction}

In his seminal paper~\cite{Majorana:1937vz}, Majorana shows that a neutral fermion, such as neutrino, can be its own anti-particle, implying then Lepton Number Violation (LNV), with an immediate consequence of neutrinoless double beta decay~\cite{Racah:1937qq,Furry:1939qr}. Often, this process is wrongly claimed to be a probe of neutrino Majorana mass, despite having been argued more than sixty years ago~\cite{Goldhaber:1959} that it could be induced by some new unknown physics. Moreover, the Left-Right (LR)~\cite{Pati:1974yy,Mohapatra:1974gc,Senjanovic:1975rk,Senjanovic:1978ev} symmetric model naturally offers new contributions, in terms of the same physics that leads to neutrino mass~\cite{Mohapatra:1979ia, Mohapatra:1980yp}. Neutrino mass is by no means special in this sense.\\

This issue is deeply related to the high-energy manifestation of lepton number violation, the hadron collider production of same-sign charged lepton pairs accompanied by jets, the so-called Keung-Senjanovic (KS)  process~\cite{Keung:1983uu}. Namely, if new physics were to induce neutrinoless double beta decay, it would have to lie at energies tantalizingly close to the LHC reach~\cite{Tello:2010am, Nemevsek:2011aa}, providing genuine hope to observe the KS process. Moreover, it would allow us to probe the seesaw mechanism as the origin of neutrino mass~\cite{Nemevsek:2012iq,Senjanovic:2016vxw,Senjanovic:2018xtu}. It would be thus imperative to untangle the physics behind neutrino mass if neutrinoless double beta decay were to be observed. This is the main scope of our work.\\

While this question would be relevant even if there was no real need for new physics, it becomes crucial because neutrino Majorana mass requires physics beyond the SM (BSM). Whether or not this new physics induces neutrinoless double beta decay, depends on its scale. One should be ready for a positive experimental result, which will require careful experimental analysis to unveil the underlying mechanism that caused it.
In short, it is impossible not to have new physics BSM along with Majorana neutrinos, and therefore it is impossible to know {\it a priori} what causes this process.\\

Before moving on, we must clarify the following
point.  In order to understand the correlation between
nature of neutrino mass and
neutrinoless double beta decay,  it is important to
know whether neutrino masses originate at the scale
$\Lambda$ above or below the characteristic momentum transfer
of this process.
It is the former case that will be studied in this work.

In other words, our analysis does not apply to
a framework with {\it soft} neutrino masses~\cite{Dvali:2016uhn},
where they originate at
a scale $\Lambda  \simeq 10^{-1}\,$eV, comparable to masses themselves.
Neutrino mass generated in this way softens above the scale in question, and
correspondingly, the connection between the neutrino mass and neutrinoless double beta decay
would require an independent analysis~\cite{private}. \\

In what follows, thus, we assume that neutrino masses and couplings are {\it hard} parameters.
With this in mind,  how can we distinguish between neutrino mass and the new physics contributions to the neutrinoless double beta decay? \\

One obvious way is by measuring the electron polarization, with an immediate conclusion: if at least one electron is Right-Handed (RH), it is new physics, period. Neutrino contribution, as we know, goes through the $W$-boson exchange so both electrons come out Left-Handed (LH). If both electrons were to be RH, it would be a boost for the Left-Right symmetric theory, which predicts the RH analog of the $W$ exchange, involving the heavy RH neutrino~\cite{Mohapatra:1979ia}, instead of the usual light one.
This is discussed in the section~\ref{UVtrue}, where it is stressed that the neutrino Majorana mass contribution can be naturally suppressed.\\

Suppose that both electrons end up being LH, what could we conclude then? Is the neutrino mass the only plausible contribution? This question is the focus of our work.
We address it from both the effective operator approach and the possible Ultra-Violet (UV) completions. While it is highly suggestive that neutrino would be the culprit, one cannot rule out other possibilities when one takes a well-defined theory as an input, - albeit admittedly not so physically natural - with new physics doing the job. More will be needed to be sure, such as establishing neutrino mass hierarchy, and/or using more than one isotope to untangle different possible contributions~\cite{Deppisch:2006hb,Simkovic:2010ka}. \\

In the next section, we use the effective operator approach to see what is going on. We shall see that new physics could be a dominant contribution, however, one cannot argue that neutrino mass is necessarily small. It should be stressed that effective operator analysis
of neutrinoless double beta decay has a long history, going back some twenty years ago (for an incomplete list of references, see~\cite{Babu:2001ex,Choi:2002bb,deGouvea:2007qla,delAguila:2012nu,Bonnet:2012kh,
Helo:2015fba,DeGouvea:2019wnq,Chen:2021rcv,Graf:2022lhj}). There have been rather general and extensive studies and here we have nothing original to say on the subject, including the issue of the resulting neutrino mass from such operators. We include this section for the sake of completeness and to be self-contained - and most importantly, to set the stage for the UV completion that is the central aspect of our work.\\

In section~\ref{UVpoor}, we discuss what might be called ``the poor man's UV completions'', where one just adds appropriate states to mediate effective operators, without worrying about the theoretical motivation. We shall see that there is always the possibility of new physics doing the job, while not contributing appreciably to neutrino mass. This should not come out as a surprise since one is fitting the phenomenological possibilities in a rather \textit{ad-hoc} manner. Again, we wish to stress that there have been such studies before, but we include this discussion here to illustrate our point and to pave the road for the analysis of a self-contained and physically motivated theory.\\

Such a theory is nicely exemplified by the original work of Pati and Salam on quark-lepton unification~\cite{Pati:1974yy}, the work that led to the idea of grand unification. This model should be considered as a true UV completion, where one makes no {\it ad-hoc} assumption and adds no states arbitrarily. The Pati-Salam (PS) model is highly constrained and, in its minimal version (at least at the renormalizable level), it is ruled out since it predicts wrong relations among down quark and charged lepton masses. This can be cured either by adding higher dimensional operators or enlarging the Higgs sector. In section~\ref{UVtrue}, we carefully analyze the situation regarding the neutrinoless double beta decay in the former case, since it is more predictive. We also comment on the latter case which allows for more freedom and thus more easily passes experimental tests. We will show that even the minimal theory can account for the neutrinoless double beta decay from new physical states and not from neutrino mass.  This is the central result of our paper which adds to the assertion that {\it per se} neutrinoless double beta decay is not automatically a probe of neutrino mass, even if electrons come out LH.\\

Finally, in the last section of the paper, we summarize our findings and offer our outlook for the future. We should stress that in this paper we only briefly touch upon theoretical ideas behind the physics in question, for a more pedagogical review see
e.g.~\cite{Senjanovic:2011zz}.

\section{Effective operator approach} \label{effective}

For simplicity and definiteness, we shall imagine here a single new physics scale $\Lambda$ responsible for neutrinoless double beta decay, hereafter denoted as $0\nu2\beta$. We will briefly comment on more general possibilities, which will be then treated with more care in the next section devoted to UV completions.

\begin{figure}[t]
\centerline{
\includegraphics[width=.65\columnwidth]{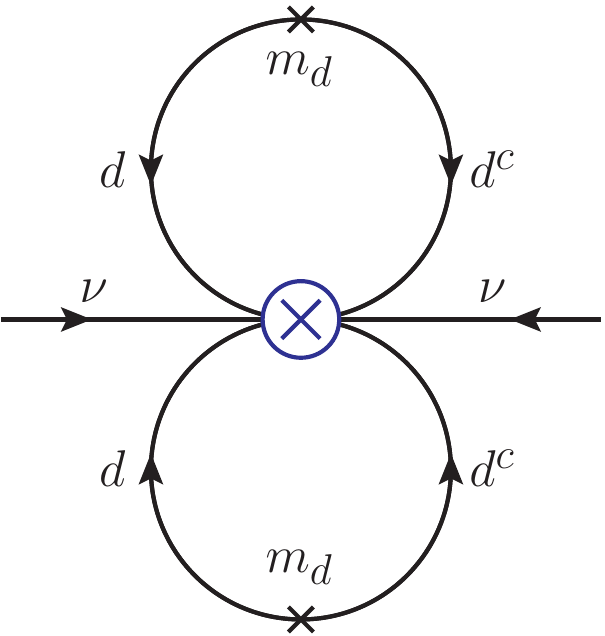}%
}
\caption{Neutrino mass in the effective operator approach.}
\label{fig:diagram1}
\end{figure}

Since $0\nu2\beta$ is a six-fermion process, just on dimensional grounds we can write a generic $d=9$ operator in a symbolic notation
\begin{equation}\label{genericO}
O_{0 \nu 2 \beta} =  A_\Lambda\,  e\, e \,u \,u \, d^c \,  d^c \,,
\end{equation}
where for simplicity we have written the hermitian conjugate operator (leaving the fermion chiralities unspecified) and defined
\begin{equation}\label{Alambda}
A_\Lambda = \frac{1}{\Lambda^5}\,.
 \end{equation}
Before plunging into our analysis, a few noteworthy comments are in order.

In the SM augmented with a neutrino Majorana mass, one has
\begin{equation}\label{SMnu}
A_\nu  \simeq G_F^2 \frac{m_\nu}{p^2}\,,
\end{equation}
where $p$ measures neutrino virtuality and is roughly $p \approx 100\, \text{MeV}$ for the nuclei relevant in question.

For $0\nu2\beta$ to be induced by new physics and not neutrino Majorana mass, we must demand
\begin{equation}\label{domin}
A_\Lambda > A_\nu
\end{equation}
Strictly speaking, one should be comparing total decay rates instead of amplitudes~\cite{Pas:1999fc, Pas:2000vn, Cirigliano:2018yza, Deppisch:2020ztt,  Cirigliano:2022oqy}. However, for our interest in physical estimates, this is a secondary task, well beyond the scope of this work.

Recently, GERDA experiment~\cite{GERDA:2020xhi} has set a strict limit  $\tau_{0\nu2\beta} \geq 10^{26} {\rm yr}$, which translates into $m_\nu \lesssim  0.2$eV. In what follows we will assume $m_\nu \simeq 0,1$eV, which corresponds to $A_\nu  \simeq 10^{-18} \text{GeV}^{-5}$, implying  $\Lambda \simeq 3 \,$TeV. This is a remarkable result since the generic scale would be tailor-made for a new hadron collider, and possibly accessible at the LHC. Both CMS and ATLAS~\cite{Aaboud:2018spl} are now actively pursuing the high energy aspect of LNV, while the low energy end is represented by several experiments dedicated to the $0\nu2\beta$.\\

A generic operator of~\eqref{genericO}, for the LH electrons, becomes then (again in a symbolic notation)
\begin{equation}\label{genericLHO}
O_{0 \nu 2 \beta} = \frac{1}{\Lambda^5} e_L\, e_L \,u_L \,u_L \,d^c_L\, d^c_L \,,
\end{equation}
which is invariant under the SM gauge symmetry if the color indices of the up quarks match the corresponding indices of the anti-down quarks. By the $SU(2)$ part of the symmetry, there must be an analogous operator with neutrinos
\begin{equation}\label{genericnu}
\frac{1}{\Lambda^5} \nu_L\, \nu_L \,d_L \,d_L \,d^c_L\, d^c_L \,.
\end{equation}
This produces neutrino Majorana mass at the two-loop level when the quark lines get closed, as shown in Fig.~\ref{fig:diagram1}. It is easy to estimate the resulting neutrino mass
\begin{equation}\label{numass}
m_\nu^M  \simeq \frac{1}{(16 \pi^2)^2}\, \frac{m_d^2}{\Lambda}\, \left (\ln \frac{\Lambda}{m_d} \right)^2\,.
\end{equation}
%
Recall that, for $0\nu2\beta$ to be observable in near future, this requires $\Lambda \simeq \,$TeV, which in turn gives
\begin{equation}\label{numassvalue}
m_\nu^M \simeq 10^{-1} \text{eV}\,,
\end{equation}
a borderline value, if we wish that neutrino mass contribution does not dominate $0\nu2\beta$. Unless there are cancellations, in this case, there is no way to argue that it is new physics, and not neutrino mass, behind this process. The operator in~\eqref{genericLHO} is not unique, though, so the resulting neutrino mass expression in~\eqref{numass} is not the only one. \\

Before we proceed, a comment is in order. We do not claim that the above estimate gives a correct value for neutrino mass. It could be smaller, of course - it could be practically zero for all that we know - but there is no way of ensuring it. \

What we are arguing, however, should not be confused with the so-called black box contribution to neutrino Majorana mass~\cite{Schechter:1980gr}, which obscures the issue. Namely, one argues that the observation of neutrinoless double beta decay demonstrates the Majorana nature of the neutrino, but that is misleading. After all, that contribution is vanishing small (argued to be on the order of $10^{-28} \text{eV}$~\cite{Duerr:2011zd}, which experimentally amounts to zero). \\

The issue of neutrino mass is a quantitative one: does neutrinoless double beta decay imply observable Majorana mass? The answer is negative since it can simply be new physics that leads to this process. If electrons come out RH, neutrino Majorana mass can be in principle as small as one wishes since, in this case, it is surely different physics that causes neutrinoless double beta decay. If electrons come out LH, the situation is more subtle and our point is that the natural expectation is non-negligible neutrino Majorana mass, incomparably larger than the naive black box argument.\\

Regarding the possible effective operators: The list is too long to be displayed here, and can be found in some of the
papers~\cite{Babu:2001ex,Choi:2002bb,deGouvea:2007qla,delAguila:2012nu,Bonnet:2012kh,DeGouvea:2019wnq,Chen:2021rcv}. In what follows we rather stick to the scalar operators for simplicity since they suffice for our task.

There are three different scalar-type operators with LH electrons -- depending on whether up and down quarks are LH or RH -- invariant under the SM gauge symmetry. We write explicitly $SU(2)$ and Lorentz structures, but suppress the color indices for the sake of notational simplicity -- it is a trivial exercise to restore them when needed. In our notation, $\ell_L$ and $q_L$ are LH lepton and quark doublets, $f^c_{L,R} \equiv C (\overline f_{R,L})^T$ and $C$ is the usual Dirac conjugation matrix, and the doublet $q^c_L$ requires the asymmetric matrix $i \sigma_2$ to transform correctly under $SU(2)_L$.

Here is the list
\begin{align}
& O_{0 \nu 2 \beta} ^{(1)}= \frac{1}{\Lambda^5} (\ell_L^T C \,  \sigma_2  \vec \sigma\, \ell_L) (q^{cT}_R C \,  \sigma_2  \vec \sigma\, q^c_R)  (u_R^T C  u_R)\label{operator1} \\
& O_{0 \nu 2 \beta} ^{(2)}= \frac{1}{\Lambda^5} (\ell_L^T C \,  \sigma_2  q_L) (\ell_L^T C d^c_L)   (u_R^T C  q^c_R)  \label{operator2} \\
& O_{0 \nu 2 \beta} ^{(3)}= \frac{1}{\Lambda^5} (\ell_L^T C \,  \sigma_2  q_L) (\ell_L^T C \,  \sigma_2  q_L) (d^{cT}_L C  d^c_L) \label{operator3}
\end{align}
It is easy to see that several operators produce neutrino mass analogous to the one in~\eqref{numass}, with different dependence on up and down quark masses.
At the level of the qualitative analysis presented in this work, however, this matters very little, and we take $m_d \simeq m_u \equiv m_q$ in what follows.\\

The bottom line regarding the effective operator approach, as we showed above, is that the situation is borderline, inconclusive as to what is causing the $0\nu2\beta$ and so we turn now to the explicit renormalizable models. Our task is facilitated by the fact that the form of the operators suggests the type of UV completion: each bi-fermion combination fixes the quantum numbers of the scalar mediators.

\section{(Poor man's) UV completion} \label{UVpoor}

It is straightforward to come up with an \textit{ad-hoc} UV completion of the above operators, a simple-minded model building that we coined ``poor man's'', for the lack of a better name. Despite the lack of theoretical structure in this approach, one still ends up with important phenomenological predictions of quite light mediators of the above operators, potentially accessible at the LHC and reachable at the next hadron collider. Moreover, the analysis presented here serves to investigate in the next section the scenario based on the PS model.\\

It is noteworthy that the UV completion is dictated by the form of the effective operators above. The messengers responsible for these operators must be either di-quarks and lepto-quarks or di-quarks and di-leptons. In what follows, we shall illustrate the possibilities with a couple of examples - the reader can easily construct variations on the theme.

\subsection{Model I}

\begin{figure}[t]
\centerline{
\includegraphics[width=.65\columnwidth]{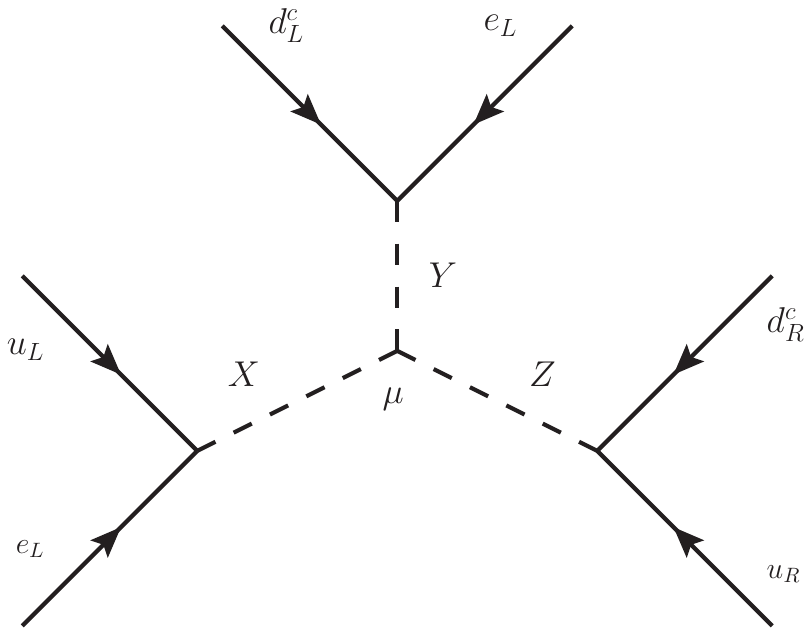}%
}
\caption{$0\nu2\beta$ in Model I, due to exchange of scalar mediators $X, Y, Z$.}
\label{fig:diagram2}
\end{figure}

We wish to be as general as possible, and so we complete first the operator given in~\eqref{operator2}, which requires the maximum number of three different scalar mediators
\begin{align}\label{UV}
& y_x  X \, q_L ^T i\sigma_2  C \ell_L  + y_y Y^T i \sigma_2 \, d^{c T}_L  C \ell_L  +  y_z Z^T i \sigma_2 \, u^T_R C q^c_R \nonumber \\
&+  \mu \, X\,Y^T i \sigma_2 \,Z + \text{h.c.}\,,
\end{align}
where $X$ is a $SU(3)_c$ triplet, $SU(2)_L$ singlet,  $Y$ is a $SU(3)_c$ triplet, $SU(2)_L$ doublet, and $Z$ is a $SU(3)_c$ sextet, $SU(2)_L$ doublet, with the $B-L$ quantum numbers 2/3, 4/3 and 0, respectively. The $B-L$ symmetry is then broken by the $d=3$ soft term $\mu$. In what follows we suppress the color indices. Their exchange produces the $0\nu2\beta$ and the neutrino mass, as shown in Fig.~\ref{fig:diagram2}.

It is evident that now there is more freedom, having four, in principle different, scales, $m_X$, $m_Y$, $m_Z$, and $\mu$. This, of course, tells us that the operator analysis with just one scale is inconclusive. In order to be as conservative as possible in our analysis, hereafter we take $y_x \simeq y_y \simeq y_z \simeq 1$. Notice that smaller Yukawa couplings will imply lower limits on the masses of scalar messengers, making these states more accessible experimentally.  \\

We can make a connection with the effective operator analysis through a simple identification
\begin{equation}\label{UVident}
A_\Lambda = \frac{\mu }{ m_X^2 \,m_Y^2\,  m_Z^2}\,.
\end{equation}
Using~\eqref{domin} and~\eqref{UVident} implies
\begin{equation}
m_\nu \, m_X^2 \, m_Y^2 \, m_Z^2  \,  G_F^2 \leq \mu \, p^2 \,.
\end{equation}
On the other hand, one has for neutrino mass
\begin{equation}\label{numassagain}
m_\nu^M \simeq \left (\frac {1}{16 \pi^2} \right)^2 \mu \, \frac {m_q^2}{m_H^2}\,,
\end{equation}
where $H$ denotes the heaviest of the scalar mediators.
These two equations give, in turn
\begin{equation}\label{mYmZ}
\frac {m_X\,m_Y \, m_Z}{m_H} \lesssim 16 \pi^2 G_F^{-1} \frac{p}{m_q} \,.
\end{equation}
where, as we said, $p \simeq 100\, \text{MeV}$.\\

The precise quantitative results depend on two possible scenarios: a) $ m_X \gg m_Y, m_Z$, and b) $ m_X \simeq  m_Y \simeq  m_Z$, which we now discuss in detail.
Strictly speaking, one has two other choices: $ m_Y \gg m_X, m_Z$, and $ m_Z \gg m_X, m_Y$. However, the numerical analysis does not depend on this since the three states $X$, $Y$, and $Z$ enter symmetrically in the $0\nu2\beta$ and the loop diagrams for induced neutrino mass. In other words, these two possibilities go under case a) - hence, only two different cases.\\
\begin{enumerate}

\item[a)]

In this case, one immediately has from~\eqref{mYmZ}
\begin{equation}\label{mYmZ2}
m_Y \, m_Z \lesssim 10^{8} \,\text{GeV}^2 \,.
\end{equation}
From the lepto-quark and di-quark searches, one has a direct limit $m_{X,Y,Z} \gtrsim 2 \text{TeV}$. Under the assumption $m_X \gg m_Y, m_Z$, this implies roughly $m_X \gtrsim 10 \text{TeV}$.

We must now distinguish two possible scenarios: either $m_Y \simeq m_Z$ or $m_Y \gg m_Z$.

In the former case, one gets $m_X, m_Y \lesssim 10^4 \text{GeV}$, which would guarantee their discovery at the next hadron collider.
Of course, in case this limit is saturated, the assumption $m_X \gg m_Y, m_Z$ would make $m_X \gtrsim 100 \text{TeV}$, outside the experimental reach in the near future.

The latter case is more interesting since then the hierarchy implies that one of the scalar messenger masses lies close to the experimental limit (order TeV), thus potentially observable even at the LHC. As we remarked before, the situation is completely symmetric in the exchange of $X, Y, Z$ states - the only thing that changes are the detailed properties of the heavier (lighter) states. In this sense, a more detailed analysis, beyond the scope of this work, may be called for. \\

While all this is necessary for the exchange of the scalars in question to dominate $0\nu2\beta$ over the neutrino contribution, it is not sufficient.  One must also make sure that the neutrino mass is sufficiently small. From~\eqref{domin},~\eqref{UVident}, and~\eqref{mYmZ}, one has
\begin{equation}\label{mumX}
\frac{\mu}{m_X^2} \lesssim 10^{-2} \text{GeV}^{-1} \,.
\end{equation}
which, using~\eqref{numassagain}, gives $m_\nu \lesssim 10^{-2} \,\text{eV}$, which is safely below the neutrino mass limits, and below the reach of this generation $0\nu2\beta$ experiments. This guarantees the self-consistency of new physics responsible for $0 \nu 2 \beta$ if this process were to be observed in the near future.\\

\item[b)]
In this case, $ m_X \simeq  m_Y \simeq  m_Z$. The same analysis as the one above obviously produces the same limit
given in~\eqref{mYmZ}, implying that all the masses lie below  $m_X, m_Y, m_Z \lesssim 10^4 \text{GeV}$. If not discovered at the LHC, all three scalar lepto-quark and di-quark states would be accessible at the next hadron collider.\\

\end{enumerate}
If we relax our assumptions of Yukawa couplings being of order one, things are even more interesting since the masses of the scalar mediators $X, Y, Z$ become smaller, and thus potentially more observable: the limits simply scale down as $\sqrt y$.\\

The detailed predictions depend on the relations between the masses of the scalar mediators that provide the UV completion of the effective operator~\eqref{operator3}, but at least one of the lepto-quarks of di-quarks must lie below 10 TeV, guaranteeing its observability at the next hadron collider. This is obtained for large values of Yukawa couplings, of order one; for smaller values, such a state becomes accessible even at the LHC - providing that the $0\nu2\beta$ is mediated by these particles.

\subsection{Model II}

Strictly speaking, one does not need yet another poor man's UV completion, but this particular model version is a prototype of what happens in the PS model. It provides the
completion of the effective operator in~\eqref{operator3}, and can be achieved by the exchange of only two scalar mediators through the following interaction
\begin{equation}\label{UV2}
 y_{q \ell}   \,q_L ^T C \sigma_2  \Delta_{q \ell}\, \ell_L   + y_{d d}  \Delta_{d d}\, d^{c T}_L  C d^c_L + \mu \, \Delta_{q \ell}^2 \,\Delta_{d d} + h.c.\,,
 \end{equation}
where $\Delta_{q \ell}$ and $\Delta_{d d}$ are color triplet and color sextet, weak singlet scalars, respectively, with $B-L=2/3$. The source of $B-L$ breaking is the soft $\mu$ term, as illustrated in Fig.~\ref{fig:diagram2}.

In the PS model, $\Delta_{q \ell}$ becomes a weak triplet, but this is a minor point of no real physical consequence, and thus it is instructive to see what is going on here. Our task is made simple by the fact that it is already contained in the discussion of Model I, with having $X,Y$ bosons merge into a single state, denoted here as $\Delta_{q \ell}$. In other words, it is just a case of $m_X = m_Y$.  \\

It is thus straightforward to translate the results of Model I for this case. Once again, one has~\eqref{mYmZ}, where one simply replaces $X,Y$ with $\Delta_{q \ell},\Delta_{d d}$, with the same numerical result as in~\eqref{mYmZ2}. We have either the hierarchical or degenerate mass situation, with an outcome similar to the previous case. Once again, either $\Delta_{q \ell}$ or $\Delta_{d d}$ must lie below 10 TeV, and possibly close to their lower limits on the order of TeV since one has
\begin{equation}\label{deltamasslimits}
m_{q \ell} m_{dd} \lesssim 10^8 {\rm GeV}^2\,.
\end{equation}
The value for the neutrinoless double beta decay amplitude depends then also on the value of $\mu$ in \eqref{UV2} since we have
\begin{equation}\label{amplitudeII}
A_\Lambda = \frac{\mu}{m_{ql}^4 m_{dd}^2} \simeq 10^{-18} {\rm GeV}^{-5}\,.
 \end{equation}
Since $\mu$ is arbitrary, this can be always satisfied.\\

What happens when we bring other generations into the game? If there was an analogous of~\eqref{UV} for the second generation, there would be a direct tree-level exchange of $\Delta_{d d}$ leading to $K - \bar K$ mixing, implying, in turn, a lower limit $m_{\Delta_{d d}} \gtrsim 10^6 \text{GeV}$. On the other hand, the exchange of $\Delta_{q \ell}$ is easily seen to produce  $D^0 \to \mu \bar e$ decay, and the resulting limit $m_{\Delta_{q \ell}} \gtrsim 10^5 \text{GeV}$.
This would effectively kill the exchange of these mediators as a source of $0\nu2\beta$. The way out would be to
have extra generations to couple weakly to them, which can be made technically natural by the emergence of global generational symmetry in the decoupling limit. \\

In short, once again, even if both electrons in the $0\nu2\beta$ end up being left-handed, this process can still originate from new physics and a negligible Majorana neutrino mass. It is not the most natural of possibilities, but it is not inconsistent either. The crucial point is that it is potentially testable at the LHC or the next hadron collider.

\section{UV completion: Pati-Salam model} \label{UVtrue}

One natural example to embed  \textit{ad hoc} scenarios with lepto-quark and di-quark scalars is the Pati-Salam quark-lepton unification. It is based on the $SU(2)_L \times SU(2)_R \times SU(4)_C$ gauge symmetry, augmented with the discrete LR symmetry which can be either a generalized parity or charge conjugation. Hereafter, we chose parity as in the original version of the theory, however, similar results emerge in the case of charge conjugation.\\

Quarks and leptons belong to the fundamental representations $(2,1,4)$ and $(1,2,4)$, where the numbers in the brackets denote the representation content under $SU(2)_L$, $SU(2)_R$ and $SU(4)_C$ gauge groups, respectively. Explicitly
\begin{equation}\label{q&l}
 f_{L,R} = \left(\begin{array}{c c}
u & \nu  \\
d & e
\end{array}\right)_{L,R}\,,
 \end{equation}
where we suppress the color index on quarks. Leptons are simply the fourth color, to be broken at the PS scale $M_{PS}$.\\

\paragraph{\bf{Yukawa sector}} The quark-lepton unification {\it per se} does not imply LNV, at least not from the exchange of the new colored lepto-quark gauge bosons $X_{PS}$ - the outcome depends on the symmetry breaking and the choice of the Higgs sector of the theory. A popular choice corresponds to the seesaw mechanism for neutrino mass, which requires the following representations responsible for the Majorana masses of the RH neutrinos
\begin{equation}\label{Delts}
\Delta_L = (3, 1, \overline{10})  \,, \,\,\,\,\, \Delta_R = (1, 3, \overline{10})\,,
\end{equation}
and the $SU(2)_L \times SU(2)_R$ complex bi-doublet (which contains the usual SM Higgs)
\begin{equation}\label{Phi}
\Phi = (2, 2, 1)\, ,
\end{equation}
responsible for the masses of the charged fermions and the Dirac neutrino mass matrix.
The symmetry-breaking pattern is given by
\begin{equation}\label{pattern}
G_{PS} \xrightarrow[\text{}]{\langle \Delta_R \rangle} G_{SM}   \xrightarrow[\text{}]{\langle \Phi \rangle}    U(1)_{em} \times SU(3)_C\,.
\end{equation}
The Yukawa couplings are given by
\begin{align}\label{Yukawa}
 \mathcal {L}_Y & = f_{L}^T \, C \, i \sigma_2 Y_\Delta \, \Delta_L \, f_L + f_{R}^T \, C \, i \sigma_2 Y_\Delta \,\Delta_R \, f_R \nonumber \\
&+ \bar{f_{L}} (Y_\Phi \,\Phi  + Y_{\tilde \Phi} \, \tilde \Phi) f_R + \text{h.c.}  \,,
\end{align}
where $\tilde \Phi = \sigma_2 \Phi^* \sigma_2 = (2, 2, 1)$.
Under generalized parity, the fields transform as
\begin{equation}\label{parity}
f_{L}  \to  f_{R}, \,\,\,\,\,  \Delta_L \to \Delta_R,\,\,\,\, \Phi \to \Phi^\dagger,\,\,\,\,  \tilde \Phi \to \tilde\Phi^\dagger  \,,
\end{equation}
which implies the same Yukawas for $\Delta_L$ and $\Delta_R$, used in~\eqref{Yukawa}. Moreover,
\begin{equation}\label{Yphi}
Y_\Phi^\dagger = Y_\Phi, \,\,\,\, Y_{\tilde \Phi} ^\dagger = Y_{\tilde \Phi}, \,\,\,\,     Y_\Delta^T = Y_\Delta\,.
\end{equation}
It is the neutral component of the $\Delta_R$ field whose vev $v_R$ breaks the $SU(2)_R$ symmetry and gives a large mass to the RH neutrino, paving the way for the seesaw mechanism.    \\

\paragraph{\bf{Charged fermion masses}} The essential point is that the bi-doublet $\Phi (2, 2, 1)$ is a singlet under $SU(4)_C$ group and so it gives the same masses to charged leptons and down quarks. Moreover, the neutrino Dirac mass matrix is then equal to the up-quark mass matrix
\begin{equation}\label{diracmass}
m_e = m_d\,; \,\,\,\,\, \,\,m_D = m_u\,,
\end{equation}
This is not realistic, of course. If one sticks to the minimal version of the theory, one must then include the higher dimensional (d=6) operators of the type
\begin{equation}\label{d=6yukawa}
 \frac{1}{\Lambda^2} \bar f_L  \, \Phi  \,\Delta_R^\dagger \,\Delta_R  \,f_R + (R \to L) \,,
\end{equation}
where, for simplicity and transparency, we omit the flavor structure. There are also analogous terms with $\Phi \to \tilde \Phi$ and
$\Delta_R \to \Delta_R^\dagger$.

In turn, through $\langle \Delta_R \rangle \simeq M_{PS}$, the above interaction induces corrections to $M_e$ and $M_D$ on the order of
$(M_{PS}/ \Lambda)^2$ ,
while keeping quark mass matrices intact. The point is that $\langle \Delta_R \rangle $ lies in the direction orthogonal to color - it only breaks the original $SU(4)_C$ quark-lepton symmetry and thus acts on the leptonic degrees of freedom. This automatically splits the equality of down quark and charged lepton mass matrices.

In order to account for the $b-\tau$ system, one needs $Y_b \lesssim 10^{-2}$ at the PS unification scale,  which leads to a constraint $\Lambda \lesssim 10 M_{PS}$.
This has been studied in the context of grand unification, see e.g.~\cite{Dorsner:2006fx}, but the effect is the same, as long as the Pati-Salam scale is close to the grand unification one. The crucial point is that the main impact comes from the strong coupling running of quark masses.
This is sufficient to guarantee the consistency of the expansion in $M_{PS}/\Lambda$, under the proviso  $\Lambda \simeq 10 M_{PS}$, assumed hereafter.\\

\paragraph{\bf{Neutrino mass}} The physical meaning of $Y_\Delta$ is provided by the emerging seesaw picture from the high-scale symmetry breaking. By defining $N_L\equiv C \bar{\nu}_R^T$, one gets
\begin{equation}
M_N=Y_\Delta^* \langle \Delta_R \rangle \,.
\end{equation}
In other words, $Y_\Delta$ provides the large masses of the RH neutrinos.
Then, the seesaw mechanism gives
\begin{equation}\label{numass_seesaw}
M_\nu \simeq - M_D^T M_N^{-1} M_D \,.
\end{equation}
The case of the third generation is rather interesting since the top Yukawa coupling is large and thus the higher dimensional contribution from~\eqref{d=6yukawa} can be neglected,
\begin{equation}\label{mD3}
m_{D_3} \simeq m_t\,.
\end{equation}
From \eqref{numass_seesaw}, the smallness of neutrino mass, $m_\nu \lesssim 1 \text{eV}$, leads to  $m_{N_3} \gtrsim 10 ^{13} \,\text{GeV}$.  From the perturbativity argument $Y_\Delta \lesssim 1$, this implies a lower limit on the unification scale
\begin{equation}\label{PSscale}
M_{PS} \simeq \langle \Delta_R \rangle \gtrsim10^{13} \text{GeV}\,.
\end{equation}
This fits nicely -- and is an independent argument -- with the above requirement that the PS scale is huge, similar to the case of grand unification.  \\

\paragraph{\bf{Neutrinoless double beta decay}}

Using the decomposition of the symmetric representation $10$ of $SU(4)_C$ under the usual $SU(3)_C$: $10 = 6 + \bar 3 +1$, we see that besides the $B - L = \pm 2$ weak triplets, we have the $B-L = \pm 2/3$ color triplets and sextets, the quantum numbers of our fields $\Delta_{q \ell}$ and $\Delta_{d d}$ in the notation of the previous section. However, instead of being a weak singlet as in the previous section,  the field $\vec{\Delta}_{q \ell}$ now becomes a triplet, whose component $\Delta_{q \ell}$ with $T_{3L} = 0$ plays the role of a mediator in \eqref{UV2}. in In particular, our fields $\vec{\Delta}_{q \ell}$ and $\Delta_{d d}$ are defined by
\begin{align}\label{ABexplicit}
& \vec{\Delta}_{q \ell}(\bar 3_C)  \in \Delta_L^\dagger, \,\,\,\,T_{3 L} (\Delta_{q \ell})  =  (0, \pm 1), \,\, \,\,\,\,\,Y/2 (\Delta_{q \ell})=  - 1/3   \nonumber  \\
& \Delta_{d d} (6_C)  \in \Delta_R^\dagger, \,\,\,\,\,\,\,\,\,\,T_{3 R} (\Delta_{d d}) = -1,\,\,\,\,\,\,\,\,  Y/2 (\Delta_{d d})= -  2/3\,.
\end{align}
Notice that the whole LH triplet  $\vec{\Delta}_{q \ell}$  must be light due to the SM symmetry (this will be important in the study of the unification constraints).\\

The $\mu$ term of~\eqref{UV}  that gives the cubic interaction $\vec{\Delta}_{q \ell} \cdot  \vec{\Delta}_{q \ell}\, \Delta_{d d}^*$ results from the $SU(2)_R$ breaking, $\mu = \alpha \langle \Delta_R \rangle = \alpha M_{PS}$, where  $\alpha$ denotes the the quartic coupling
\begin{equation}\label{alphacoupling}
\alpha Tr (\Delta_L \Delta_L) Tr (\Delta_R^\dagger \Delta_R^\dagger)\,.
\end{equation}
Using $\alpha \lesssim 1$ from perturbativity requirements, one has $\mu \lesssim M_{PS}$.
In turn,  for the $0\nu2\beta$ amplitude to be potentially observable, \eqref{amplitudeII} requires
\begin{equation}\label{amplitudePS}
A_{PS} \lesssim \frac{M_{PS}}{m_{ql}^4 m_{dd}^2} \simeq 10^{-18} {\rm GeV}^{-5}\,.
 \end{equation}
At this point, the unification scale $M_{PS}$ is undetermined. We only assume that it lies sufficiently below the Planck scale in order to stay away from gravity becoming strong and thus losing the calculability. In what follows, we then take $M_{PS} \lesssim 10^{16}$ GeV and thus we obtain $m_{ql}^4 m_{dd}^2 \gtrsim 10^{34} \text{GeV}^6$. On the other hand, from the discussion below \eqref{amplitudeII}, from flavor conservation one has the lower limits on these masses  $m_{\Delta_{d d}} \gtrsim 10^6 \text{GeV}$ and $m_{\Delta_{q \ell}} \gtrsim 10^5 \text{GeV}$. Clearly, the scalar states of the Pati-Salam theory can dominate neutrinoless double beta decay over neutrino exchange, without running into the problem of flavor. This is a central result of our paper. \\

We should bear in mind, though, that the Pati-Salam scale could be lower and of course, the coupling $\alpha$ could be small, and thus the $\mu$ term itself could be substantially lower. Moreover, either (or both) of these scalar states might be light and accessible to experiment -- any of these possibilities would bring back the issue of flavor. However, it can be shown that the Yukawa structure of the theory is rich enough to avoid the flavor problem even in the extreme situation of both $\Delta_{q \ell}$ and $\Delta_{d d}$ lying close to TeV energies. \\

In order to see that, we start with
the relevant part of the Yukawa Lagrangian, taken from~\eqref{Yukawa}
\begin{equation}
\mathcal{L}_Y \supset f_L^{0\, T} Y_{\Delta} \Delta_L  f_L^0 + f_R^{0\, T} Y_{\Delta} \Delta_R f_R^0\,.
\end{equation}
where the upper script $0$ denotes the original weak eigenstates before the diagonalization of mass matrices, and as discussed above, the LH and RH Yukawa couplings are the same due to parity.

Since we are interested in the flavor structure in what follows, for the sake of simplicity and transparency, we shall suppress the $SU(2)$ and Lorentz indices, so that we can simply write
\begin{equation}\label{Lag_PS_flavor}
\mathcal{L}_Y \supset  u_L^{0\, T} Y_\Delta e_L^0 \Delta_{q \ell} + d_R^{0\, T} Y_\Delta d_R^0 \Delta_{d d} + h.c. \,,
\end{equation}
where by $\Delta_{q \ell}$ we denote, as originally in \eqref{UV2}, the component of the triplet $\vec{\Delta}_{q \ell}$ with $T_{3L}= 0$.
The charged fermion mass matrices are bi-diagonalized via the unitary transformations $F_{L,R}$
\begin{equation}
f_{L,R}^0= F_{L,R} f_{L,R}\,,
\end{equation}
so that~\eqref{Lag_PS_flavor} leads to
\begin{equation}
\mathcal{L}_Y= u_L^{T} Y_{q \ell} e_L \Delta_{q \ell} + d_R^{T} Y_{d d} d_R \Delta_{d d} + h.c. \,,
\end{equation}
with
\begin{align}
& Y_{q \ell}= U_L^T Y_\Delta E_L  \label{yDA}   \\
& Y_{d d}= D_R^T Y_\Delta D_R  \label{yDB}\,.
\end{align}
Recall that, to have neutrinoless double beta decay dominated by the exchange of new states (and not by neutrino mass), the associated Yukawa couplings need to be sufficiently large:  $Y_{q \ell} \simeq Y_{d d} \simeq 1$. In turn, this implies $Y_\Delta \simeq 1$ or, in other words, $M_N \simeq M_{PS}$. It fits nicely with~\eqref{mD3} and a large Pati-Salam scale.\\

Kaon mixing constrains the part of Lagrangian proportional to $Y_{d d}$: either $\Delta_{d d}$ is extremely heavy, or $Y_{dd}$ is flavor diagonal. The former possibility clashes with the requirement that neutrinoless beta decay is triggered by the exchange of $\Delta_{q \ell}$ and $\Delta_{d d}$ states, and so we opt for the latter. For simplicity and illustration, we choose
\begin{equation}\label{yBdiag}
Y_{dd} \simeq \text{diag} ( 1, 1, 1 )\,,
\end{equation}
which ensures no flavor violation and $\Delta_{d d}$ can be as light as its direct experimental limit $m_{\Delta_{dd}}\gtrsim \mathcal{O}(1)$TeV. Notice that $Y_\Delta$ from~\eqref{yDB} in general violates flavor. \\

From~\eqref{yDA} and~\eqref{yDB}, one immediately gets
\begin{align}
Y_{q \ell} & = U_L^T D_R^* Y_{d d} D_R^\dagger E_L = \nonumber \\
& \simeq V_1 V_2  \label{yAintermedio}\,,
\end{align}
where $V_1 = U_L^T D_R^*$ and $V_2 = D_R^\dagger E_L$. Since there is no connection between $M_e$ and $M_d$ due to \eqref{d=6yukawa}, the matrix $V_2$ is completely arbitrary. On the other hand, the theory keeps the memory of the original parity symmetry implying $D_L \simeq D_R$~\cite{Senjanovic:2014pva,Senjanovic:2015yea}, so that $V_1 \simeq V_{CKM}^*$. In order to avoid the decay $D^0 \to \mu \bar e$, it is sufficient that the $\Delta_{q \ell}$ interaction takes the form.
\begin{equation}\label{Yql}
Y_{q \ell} \simeq \left(
\begin{array}{ccc}
1 & 0 & 0 \\
0 & 0 & 1 \\
0 & 1 & 0
\end{array}
\right )\,,
\end{equation}
which, due to the arbitrary nature of $V_2$,  can be obviously achieved with the choice $V_2 = V_{CKM}^T Y_{q \ell}$.
Thus, it is perfectly consistent to have both $\Delta_{q \ell}$ and $\Delta_{d d}$ interactions in accord with limits on flavor violation, so that their masses can be as low as the direct experimental limits of the order of TeV. \\

In conclusion, at least on phenomenological grounds, it is perfectly possible within this theory to have neutrinoless double beta decay induced dominantly by new physical states, instead of neutrino Majorana mass. For the sake of illustration, we give two benchmarks (limiting scenarios) examples that saturate \eqref{amplitudePS}
\begin{itemize}
\item
$m_{q \ell} \simeq m_{d d } \simeq 10 ^3$ GeV, $\mu  \gtrsim 1$ GeV\,,

\item
 $m_{q \ell} \simeq 10 ^5$ GeV, $m_{d d } \simeq 10 ^6$ GeV, $\mu  \gtrsim 10 ^{14}$ GeV\,.
\end{itemize}
In both cases, one just needs to assume $m_\nu \lesssim 0.1$ eV, for the neutrino contribution to neutrinoless double beta decay to be subdominant. Since neutrino mass is not determined by the theory, this is always possible - neutrino contribution can be made as small as needed. Needless to say, any combination of scalar masses between these extremes works as well, with an appropriate choice of $\mu$ scale, or equivalently, the coupling $\alpha$. Due to its direct experimental interest, in what follows most of our discussion will be focused on the first case of light leptoquarks and diquarks.\\

What about other possible contributions to the neutrinoless double beta decay? Besides the obvious candidates $\Delta_{q \ell}$ and $\Delta_{d d }$, there are also contributions from the RH gauge boson $W_R$ and RH neutrino N (RH analog of the usual neutrino source) and the contributions from the $SU(2)_R$ and  $SU(2)_L$ triplets $\Delta_R$ and $\Delta_L$. Except for the $\Delta_L$ contribution, the rest implies RH electrons in the final states,
which is complementary to our work. This has been amply discussed in the literature, from the original papers~\cite{Mohapatra:1979ia, Mohapatra:1980yp} on the seesaw mechanism to the connection with the hadron colliders~\cite{Tello:2010am, Nemevsek:2011aa}.
The interested reader can find more detailed and updated analysis in the following references~\cite{Mitra:2011qr,BhupalDev:2014qbx,Li:2020flq, Li:2022cuq}. \\

Once again we remind the reader, that we are after new physics contributions with purely LH electrons in the final states. A single RH electron in the final state implies automatically a new source and not neutrino Majorana mass which requires LH electrons. It is easy to see that the exchange of $\Delta_L$ triplet is suppressed by roughly $p/m_{\Delta_L}$, where $p \simeq 100$ MeV is the momentum exchange in relevant nuclei and $m_{\Delta_L} \gtrsim 800$ GeV from
experiment~\cite{Dziewit:2021pak}. There is also a contribution from $W_L$ and N from the neutrino-N mixing, which is easily seen to be negligible if N's are heavier than $p$. In case they were substantially lighter than $p$, however, this contribution would become important, since it would cancel the usual neutrino Majorana mass exchange (up to
$(m_N/p)^2$)~\cite{Blennow:2010th}. \\

In other words, for very light N, our argument of new physics being behind neutrinoless double beta decay, even with final states LH electrons, would be only strengthened. In the minimal non-renormalizable PS model, as we argue above, at least the third generation N has to be very heavy, on the order of the unification scale, eliminating this possibility. In the renormalizable version, though, N can be arbitrarily light and thus could lead to the suppression of the light neutrino exchange. For a more complete discussion of light N with the inclusion of possibly light $W_R$, see~\cite{deVries:2022nyh}.\\

What remains to be done on theoretical grounds is to check that the unification constraints allow for these examples. It turns out that there is enormous freedom in these constraints - masses can be almost completely arbitrary - which then justifies rough estimates used in the above. Before plunging into this tedious analysis, we first discuss, though, a renormalizable version of this theory.\\

\paragraph{\bf{A variation on a theme: renormalizable model}}
As we have seen, the minimal theory requires going beyond d=4 interaction, appealing to higher dimensional operators. If one asks for the theory to stay renormalizable, one must introduce an additional bi-doublet field, in the adjoint of $SU(4)$ representation
\begin{equation}\label{Phi}
\Phi' = (2, 2,15)\,,
\end{equation}
and couple it to fermions in complete analogy with the $\Phi$ couplings Yukawa couplings in~\eqref{Yukawa}.
Its vacuum expectation value, $\langle \Phi' \rangle \propto diag (1,1,1,-3)$ in the $SU(4)$ space, then leads to arbitrary charged fermion mass matrices, as was achieved before by d=6 terms in~\eqref{d=6yukawa}. Since the new Yukawa couplings need not be small, this version of the theory is less constrained and, as before, obviously satisfies the flavor bounds.\\

The thing to keep in mind, though, is that although $M_N \simeq M_{PS}$ since $Y_D$ is arbitrary, $M_N$ and thus also the unification scale can be as low as the phenomenology allows.\\

\paragraph{\bf{Unification}}
The lower bound on the Pati-Salam unification scale arises from rare kaon decays, due to the exchange of the new leptoquark gauge bosons.  The limit is subject to the unknown flavor mixing angles and by their judicious choice, it can be as low as $M_{PS} \gtrsim 10^5 \text{GeV}$~\cite{Dolan:2020doe}.
 However, in the minimal model augmented with d=6 operators, the large Dirac neutrino mass of the third generation implies a huge limit $M_{PS} \gtrsim 10^{13} \text{GeV}$. If one abandons the minimal model, however, this limit goes away and here the unification constraints play a more relevant role as we shall see.\\

The main issue in studying the unification constraints is the knowledge of relevant particle masses that participate in the running of the gauge couplings. We know the fermion and the gauge boson masses -- in the sense that the leptoquark gauge bosons lie at the unification scale -- while the masses of the scalars are the problem. They are arbitrary due to higher dimensional operators needed for the sake of charged fermion masses; these new terms effectively eliminate the sum rules that tend to follow from the renormalizable d=4 part of the Higgs potential. This enormously complicates unification analysis.\\

The same problem emerges in true grand unified models, such as $SU(5)$ or $SO(10)$. Here, as in the case of partial unification, one typically assumes the so-called extended survival principle where one argues that the scalar boson masses should take the largest possible values consistent with the symmetries of the theory in question~\cite{delAguila:1980qag,Mohapatra:1982aq}. However, this can be completely wrong since there are predictive counter-examples in which, on the contrary, some masses end up choosing the lowest possible values. In particular, the minimal SO(10) theory with small representations and higher dimensional representations predicts light states potentially accessible even at the LHC~\cite{Preda:2022izo}. A similar situation emerges in a minimal realistic version of the $SU(5)$ theory based on an adjoint fermion representation~\cite{Bajc:2006ia,Bajc:2007zf}. These examples teach us that the scalar masses are free parameters and ought to be treated as such, with the hope of their values being determined by consistency requirements and/or phenomenological considerations. This shall be followed here.\\

The starting point is the $SU(4)_C$ unification which implies the equality of B-L and QCD gauge couplings at the unification scale $M_{PS}$, $\alpha_{BL} (M_{PS}) = \alpha_3(M_{PS})$. It is well known that $(B-L)/2  = \sqrt {2/3} \,T_{BL} $, where $T_{BL}$ is a canonically normalized generator.  Next, from $ Y/ 2 = T_{3R} + (B-L)/2$ and the fact that the LH and RH $SU(2)$ gauge couplings are equal at the unification scale, one gets
\begin{equation}\label{PSunif}
\frac{1}{\alpha' (M_{PS})} = \frac{1}{\alpha_2 (M_{PS})} + \frac{2}{3} \frac{1}{\alpha_3 (M_{PS})}\,,
\end{equation}
where $\alpha'$ in the standard notation is the gauge coupling corresponding to the $U(1)_Y$ gauge group of the SM. The next step is the usual one, one evolves $\alpha'$ from the weak to the Pati-Salam scale. \\

We are now ready to determine the unification scale. The starting equations describe the running from $M_Z$ to $M_{PS}$ scales
\begin{align}\label{runninalpha}
&\frac{1}{\alpha(M_{PS})} = \frac{1}{\alpha (M_{M_Z})} + \frac{b_0}{2 \pi} \, \ln \frac{m_1}{M_Z} + \frac{b_1}{2 \pi} \, \ln \frac{m_2}{m_1} \nonumber\\
&+... \frac{b_{i-1}}{2 \pi} \, \ln \frac{m_i}{m_{i-1}} +...  + \frac{b_n}{2 \pi} \, \ln \frac{M_{PS}}{m_n}\,,
\end{align}
where $\alpha$ stands for any of the above couplings and  $m_n \gtrsim m_{n-1} ...\gtrsim m_0 = M_Z$ denotes the masses of different physical scalar states, from the heaviest to the lightest.

From~\eqref{runninalpha} and~\eqref{PSalpha} one readily obtains
\begin{equation}\label{PSscaledet}
\frac {M_{PS}}{M_Z} =  \exp \left ({\frac{2 \pi}{ b_{PS}^n \, \alpha_{PS} (M_Z)}} \right)
 \prod_{i=1}^n \left (\frac {m_i}{M_Z} \right)^{(b_{PS}^i - b_{PS}^{i-1}) /b_{PS}^n}
\end{equation}
where
\begin{equation}\label{PSalpha}
\frac{1}{\alpha_{PS}} \equiv   \frac{1}{\alpha'} - \frac{1}{\alpha_2} -  \frac{2}{3} \frac{1}{\alpha_3}\,,\,\,\,\,\,b_{PS} \equiv   \frac{2}{3} b_3 + b_2 - b'\,,
\end{equation}
and $b_3, b_2, b'$ stand for the running coefficients of the couplings $\alpha_3, \alpha_2, \alpha'$, respectively.\\

In our case, we know that for the neutrinoless double beta decay not to be dominated by neutrino mass, one needs $\vec{\Delta}_{q \ell}$ and $\Delta_{d d}$ states to satisfy \eqref{deltamasslimits}. For the sake of illustration, we choose these masses to be of order TeV since this would allow for their possible direct detection even at the LHC. This in turn fixes $\mu \simeq$ GeV which requires the quartic coupling in \eqref{alphacoupling} to be rather small.
If other scalars end up being heavy, close to the PS scale, we get $M_{PS} \simeq 10^{14} \text{GeV}$. This fits perfectly with the requirement of the realistic $b-\tau$ mass spectrum, as discussed above. Notice that this is orders of magnitude above the lower phenomenological limit, out of reach in near future experiments. The essential point is that it is perfectly consistent with the phenomenological limit~\eqref{PSscale}. \\

Increasing the values of the leptoquark and diquark scalar masses to their maximum on the order of  $10^5 - 10^6$ GeV makes little difference, not relevant at the level of the discussion employed here). Moreover, it makes no conceptual difference whatsoever and we leave this as an exercise for an interested reader.
Needless to say, saturating the upper limit values of their masses would make these states out of reach at near future colliders.\\

In the renormalizable non-minimal version the situation changes and the unification scale can be in principle arbitrary. Thus,
the question becomes whether it could be substantially lowered and brought close to experiments. The answer is negative. Its lowest possible value with light $\vec{\Delta}_{q \ell}$ and $\Delta_{d d}$ states turns out to be
\begin{equation}\label{PSscalelower}
M_{PS} \gtrsim  10^{11} \text{GeV}\,,
\end{equation}
still far from the experimental reach. It requires, besides $\vec{\Delta}_{q \ell}$ and $\Delta_{d d}$, some other states to be as light as possible, with masses on the order of TeV:
\begin{align}\label{lightstates}
&(1_c, \,Y/2= -1) \hspace{3.3em}\in \Delta_L    \nonumber \\
&\hspace{-0.6em}\left.
\begin{array}{l}
(1_c, \,Y/2=-1, -2) \vspace{0.5em}\\
(3_c, \,Y/2=-4/3) \vspace{0.5em}\\
(6_c, \,Y/2=4/3,1/3)\\
\end{array}
\right \}  \in \Delta_R \,.
\end{align}
These new colored light states, together with the $\Delta_{dd}$ generate in turn $n-\bar n$ oscillations~\cite{Mohapatra:1980qe}, in a manner similar to the neutrinoless double beta decay. Instead of the coupling $\alpha$, one has now the quartic interaction of the type $\beta \Delta_R^4$. It is clear that $\alpha$ and  $\beta$  are completely independent, and $\beta$  is arbitrary which allows having $n-\bar n$ oscillations under control (the connection between these processes in the context of what we call 'poor man's models has been studied in~\cite{Gardner:2018azu}). \\

What happens is the following. In the limit $\beta \to 0$, it is easy to see that there is an additional accidental global continuous symmetry, under which the fields transform as follows
\begin{equation}\label{accidentalsymmetry}
f_{L,R} \to e^{i \theta}  f_{L,R}\,,\,\,\,\, \Phi \to \Phi\,,\,\,\,\, \Delta_{L,R} \to e^{-2 i \theta} \Delta_{L,R}\,.
\end{equation}
This is the vector-like fermion number  (F) - no direct fermion number violation in the limit $\beta = 0$.
It is easy to see that in the process of symmetry breaking, a combination of B-L  and F remains unbroken -  readily seen to be the baryon number. Namely,  $\langle \Phi \rangle$ is invariant under F and small $\langle \Delta_L \rangle$,  induced through a tadpole~\cite{Mohapatra:1980yp}, behaves exactly as $\langle \Delta_R \rangle$ due to parity - they both preserve B.

Thus, in the limit of $\beta=0$, only L gets broken through the Majorana mass of the RH neutrino, while B remains unbroken. Whether directly through $\beta$ or indirectly through $\langle \Delta_R \rangle$, F and B-L are broken only by even units - no proton decay in the PS theory.
Notice also that due to $U(1)_F$ symmetry, small $\beta$ is self-protected, and thus its smallness is technically natural. The bottom line is that, through $\alpha$ and $\beta$ couplings, it is possible to have both $n-\bar n$ oscillations and the $0\nu2\beta$ process be induced by new physics at the same time.\\

One could argue that this whole discussion seems to be of academic interest because of the huge $M_{PS}$. It is noteworthy that this is a generic problem of the Pati-Salam model: even if one allowed arbitrary values for $m_{\Delta_{q \ell}}$ and $m_{\Delta_{d d}}$, one could at best lower $M_{PS}$ by another order of magnitude, and thus one cannot imagine probing the gauge sector of the theory in any foreseeable future. \\

Moreover, since  $m_N \simeq M_{PS}$,  RH neutrinos $N$ cannot be directly observed, and the seesaw mechanism can only be indirectly tested. However, this guarantees that the neutrinoless double beta decay cannot be induced by the $W_R$ and $N$ contribution, which simplifies our study and our conclusions. One can also ask: How generic is this prediction? More precisely, is it possible to have a low intermediate scale with LR symmetry? A positive answer would require going beyond the minimal version of the model, and the simplest possibility would be to add an $SU(2)_L \times SU(2)_R$ singled, $SU(4)_C$ adjoint Higgs representation $\Sigma = (1,1,15)$, with the unification scale $\langle \Sigma \rangle = M_{PS}$ and the intermediate one $\langle \Delta_R \rangle = M_{LR}$.
We would still need higher dimensional operators to obtain a realistic fermion mass spectrum, thus the relation~\eqref{mD3} would remain valid. From~\eqref{PSscale},  one would have  $M_{LR} \gtrsim 10^{13} \text{GeV}$,  practically implying the single-step breaking for the sake of issues studied in this work, and our results would remain intact. \\

In summary, we have proven that even in the constrained, physically motivated, Pati-Salam theory, neutrinoless double beta decay can be consistently induced by new scalars, instead of neutrino Majorana mass added to the SM. The crucial point is that one can still have left-handed electrons coming out in this process, as in the latter case.
This comes at the price of fine-tuning the scalar masses $m_{\Delta_{q \ell}}$ and $m_{\Delta_{d d}}$, thus losing some of the original motivation. However, the point is that it is both phenomenologically and theoretically feasible. One could conclude, however, that the theory leaves no additional imprint, besides the light scalar states in question.\\

\paragraph{\bf{Magnetic monopoles and domain walls}} There is more to it, though.  It is well-known~\cite{tHooft:1975psz} that
the Pati-Salam theory predicts the existence of magnetic monopoles~\cite{tHooft:1974kcl,Polyakov:1974ek}, with the mass $m_M \simeq M_{PS}$.  In addition, the spontaneous breaking of the discrete left-right symmetry leads to the existence of domain walls. Both of these topological defects would be formed by the Kibble mechanism~\cite{kibble1976topology} if the universe were to undergo a phase transition. In this case, both domain walls~\cite{Zeldovich:1974uw} and monopoles~\cite{Zeldovich:1978wj,Preskill:1979zi} potentially pose cosmological problems. While the latter could be tolerated with $M_{PS} \lesssim 10^{12}\, \text{GeV}$, the former is simply a no-go.\\

One way to cure the domain wall problem is by a small explicit bias~\cite{Vilenkin:1984ib}. In particular, it could originate from the Planck scale higher dimensional operators~\cite{Rai:1992xw}, expected to break the global symmetry in question. Such a bias~\footnote{However,  a biased discrete symmetry is disfavored
by gravity~\cite{Dvali:2018txx}, as it  implies the existence of a metastable de Sitter vacuum which is highly problematic~\cite{Dvali:2013eja, Dvali:2017eba}
due to its incompatibility with  $S$-matrix formulation of quantum gravity~\cite{Dvali:2020etd}} would suffice to get rid of domain walls, moreover, while collapsing they would also sweep away the monopoles~\cite{Dvali:1997sa}. \\

  One can also appeal to inflation~\cite{Guth:1980zm}, provided the phase transition with defect formation does not take place after the end of inflation. An inflationary scenario within (supersymmetric) LR theories was studied in~\cite{Dvali:1997uq}
incorporating the inflationary mechanism of~\cite{Dvali:1994ms}.  \\

Finally, we wish to emphasize that both domain wall~\cite{Dvali:1995cc,Dvali:1996zr} and monopole~\cite{Dvali:1995cj} problems can be simultaneously solved by an appealing possibility of symmetry non-restoration at high temperature~\cite{Weinberg:1974hy, Mohapatra:1979vr,Mohapatra:1979bt}.

\section{Conclusions and Outlook}\label{Outlook}

Neutrinoless double beta decay has emerged over the years as the central low energy lepton number violation process and is often claimed to be a probe of neutrino Majorana mass. However, as known for a long time, that is not true: new physics, necessary for providing neutrino Majorana mass, could as well do the job, and it is of great importance to disentangle different sources of this process.\\

The clearest indication of new physics behind $0\nu2\beta$ would be that at least one of the outgoing electrons is RH since in the neutrino case both electrons must come out LH. However, even if both electrons were LH, one could not exclude new physics from dominating the process, despite neutrino mass remaining a natural possibility. This has been the focus of our work, in which we have provided some simple models where neutrino mass contribution to $0\nu2\beta$ could easily be sub-dominant. Needless to say, this is not surprising when one deals with \textit{ad hoc} chosen models to do the job. \\

The real test, though, is provided by a well-defined theory, such as say
the Pati-Salam quark-lepton unification. Our study has shown that even in the very minimal version of the theory, with tight constraints, new scalar states can dominate neutrinoless double beta decay. While phenomenologically rather appealing, we should stress that this requires fine-tuning of these masses. In this sense, we should be regarded as the devil's advocates. In all honesty, we think that neutrino mass is a more natural possibility in the case of left-handed electrons, but we wish to call for caution before proclaiming a sentence on this important process.\\

It will be a hard task to determine the source of neutrinoless double beta decay, especially to prove that it is due to neutrino mass. In a sense, it will be easier to rule it out, and it would be automatic if one were to show that electrons do not emerge left-handed. On the contrary, if both electrons end up being left-handed, there may still be ways to eliminate neutrino mass as the source. One possibility is the confirmation of a normal neutrino mass hierarchy, which tends to give a too-small contribution to have $0\nu2\beta$ observable in ongoing or near-future experiments (but with time also this could be reached). Similarly, cosmology could push down the upper limit on neutrino masses, with the same conclusion. One appealing possibility, besides measuring electron polarisation as a priority, would be to compare different isotopes to untangle between possible different sources~\cite{Simkovic:2010ka}.

\section{Acknowledgements}

We wish to thank Tanmay Vachaspati for a useful correspondence regarding the cosmological production of monopoles. This work was supported in part by the Humboldt Foundation under Humboldt Professorship Award, by the Deutsche Forschungsgemeinschaft (DFG, German Research Foundation) under Germany's Excellence Strategy - EXC-2111 - 390814868, and Germany's Excellence Strategy under Excellence Cluster Origins.

\bibliographystyle{utphysmod}
\bibliography{biblio}

\providecommand{\href}[2]{#2}\begingroup\raggedright\begin{thebibliography}{10}

\bibitem{Majorana:1937vz}
E.~Majorana, ``{Teoria simmetrica dell\textquoteright{}elettrone e del
  positrone}'', \href{http://dx.doi.org/10.1007/BF02961314}{\blue Nuovo Cim.
  {\bfseries 14} (1937) 171--184}.

\bibitem{Racah:1937qq}
G.~Racah, ``{On the symmetry of particle and antiparticle}'',
  \href{http://dx.doi.org/10.1007/BF02961321}{\blue Nuovo Cim. {\bfseries 14}
  (1937) 322--328}.

\bibitem{Furry:1939qr}
W.~Furry, ``{On transition probabilities in double beta-disintegration}'',
  \href{http://dx.doi.org/10.1103/PhysRev.56.1184}{\blue Phys. Rev. {\bfseries
  56} (1939) 1184--1193}.

\bibitem{Goldhaber:1959}
M.~Goldhaber and G.~Feinberg, ``{Microscopic Tests of Symmetry Principles}'',
  \href{http://dx.doi.org/10.1073/pnas.45.8.1301}{\blue Proc. Nat. Ac. Sci. USA
  {\bfseries 45} no.~8, (1959) 1301}.

\bibitem{Pati:1974yy}
J.~Pati and A.~Salam, ``{Lepton Number as the Fourth Color}'',
  \href{http://dx.doi.org/10.1103/PhysRevD.10.275}{\blue Phys. Rev. D
  {\bfseries 10} (1974) 275--289}. [Erratum: Phys.Rev.D 11, 703--703 (1975)].

\bibitem{Mohapatra:1974gc}
R.~Mohapatra and J.~Pati, ``{A Natural Left-Right Symmetry}'',
  \href{http://dx.doi.org/10.1103/PhysRevD.11.2558}{\blue Phys. Rev. D
  {\bfseries 11} (1975) 2558}.

\bibitem{Senjanovic:1975rk}
G.~Senjanovi\'c and R.~Mohapatra, ``{Exact Left-Right Symmetry and Spontaneous
  Violation of Parity}'',
  \href{http://dx.doi.org/10.1103/PhysRevD.12.1502}{\blue Phys. Rev. D
  {\bfseries 12} (1975) 1502}.

\bibitem{Senjanovic:1978ev}
G.~Senjanovi\'c, ``{Spontaneous Breakdown of Parity in a Class of Gauge
  Theories}'', \href{http://dx.doi.org/10.1016/0550-3213(79)90604-7}{\blue
  Nucl. Phys. B {\bfseries 153} (1979) 334}.

\bibitem{Mohapatra:1979ia}
R.~Mohapatra and G.~Senjanovi\'c, ``{Neutrino Mass and Spontaneous Parity
  Nonconservation}'', \href{http://dx.doi.org/10.1103/PhysRevLett.44.912}{\blue
  Phys. Rev. Lett. {\bfseries 44} (1980) 912}.

\bibitem{Mohapatra:1980yp}
R.~N. Mohapatra and G.~Senjanovi\'c, ``{Neutrino Masses and Mixings in Gauge
  Models with Spontaneous Parity Violation}'',
  \href{http://dx.doi.org/10.1103/PhysRevD.23.165}{\blue Phys. Rev. D
  {\bfseries 23} (1981) 165}.

\bibitem{Keung:1983uu}
W.-Y. Keung and G.~Senjanovi\'c, ``{Majorana Neutrinos and the Production of
  the Right-handed Charged Gauge Boson}'',
  \href{http://dx.doi.org/10.1103/PhysRevLett.50.1427}{\blue Phys. Rev. Lett.
  {\bfseries 50} (1983) 1427}.

\bibitem{Tello:2010am}
V.~Tello, M.~Nemev\v{s}ek, F.~Nesti, G.~Senjanovi\'c, and F.~Vissani,
  ``{Left-Right Symmetry: from LHC to Neutrinoless Double Beta Decay}'',
  \href{http://dx.doi.org/10.1103/PhysRevLett.106.151801}{\blue Phys. Rev.
  Lett. {\bfseries 106} (2011) 151801},
  \href{http://arxiv.org/abs/1011.3522}{{\blue arXiv:1011.3522 [hep-ph]}}.

\bibitem{Nemevsek:2011aa}
M.~Nemev\v{s}ek, F.~Nesti, G.~Senjanovi\'c, and V.~Tello, ``{Neutrinoless
  Double Beta Decay: Low Left-Right Symmetry Scale?}'',
  \href{http://arxiv.org/abs/1112.3061}{{\blue arXiv:1112.3061 [hep-ph]}}.

\bibitem{Nemevsek:2012iq}
M.~Nemev\v{s}ek, G.~Senjanovi\'c, and V.~Tello, ``{Connecting Dirac and
  Majorana Neutrino Mass Matrices in the Minimal Left-Right Symmetric Model}'',
  \href{http://dx.doi.org/10.1103/PhysRevLett.110.151802}{\blue Phys. Rev.
  Lett. {\bfseries 110} no.~15, (2013) 151802},
  \href{http://arxiv.org/abs/1211.2837}{{\blue arXiv:1211.2837 [hep-ph]}}.

\bibitem{Senjanovic:2016vxw}
G.~Senjanovi\'c and V.~Tello, ``{Probing Seesaw with Parity Restoration}'',
  \href{http://dx.doi.org/10.1103/PhysRevLett.119.201803}{\blue Phys. Rev.
  Lett. {\bfseries 119} no.~20, (2017) 201803},
  \href{http://arxiv.org/abs/1612.05503}{{\blue arXiv:1612.05503 [hep-ph]}}.

\bibitem{Senjanovic:2018xtu}
G.~Senjanovi\'c and V.~Tello, ``{Disentangling the seesaw mechanism in the
  minimal left-right symmetric model}'',
  \href{http://dx.doi.org/10.1103/PhysRevD.100.115031}{\blue Phys. Rev. D
  {\bfseries 100} no.~11, (2019) 115031},
  \href{http://arxiv.org/abs/1812.03790}{{\blue arXiv:1812.03790 [hep-ph]}}.

\bibitem{Dvali:2016uhn}
G.~Dvali and L.~Funcke, ``{Small neutrino masses from gravitational
  \ensuremath{\theta}-term}'',
  \href{http://dx.doi.org/10.1103/PhysRevD.93.113002}{\blue Phys. Rev. D
  {\bfseries 93} no.~11, (2016) 113002},
  \href{http://arxiv.org/abs/1602.03191}{{\blue arXiv:1602.03191 [hep-ph]}}.

\bibitem{private}
N.~Arkani-Hamed and L.~Funcke {private communication} .

\bibitem{Deppisch:2006hb}
F.~Deppisch and H.~Pas, ``{Pinning down the mechanism of neutrinoless double
  beta decay with measurements in different nuclei}'',
  \href{http://dx.doi.org/10.1103/PhysRevLett.98.232501}{\blue Phys. Rev. Lett.
  {\bfseries 98} (2007) 232501},
  \href{http://arxiv.org/abs/hep-ph/0612165}{{\blue arXiv:hep-ph/0612165}}.

\bibitem{Simkovic:2010ka}
F.~Simkovic, J.~Vergados, and A.~Faessler, ``{Few active mechanisms of the
  neutrinoless double beta-decay and effective mass of Majorana neutrinos}'',
  \href{http://dx.doi.org/10.1103/PhysRevD.82.113015}{\blue Phys. Rev. D
  {\bfseries 82} (2010) 113015}, \href{http://arxiv.org/abs/1006.0571}{{\blue
  arXiv:1006.0571 [hep-ph]}}.

\bibitem{Babu:2001ex}
K.~S. Babu and C.~N. Leung, ``{Classification of effective neutrino mass
  operators}'', \href{http://dx.doi.org/10.1016/S0550-3213(01)00504-1}{\blue
  Nucl. Phys. B {\bfseries 619} (2001) 667--689},
  \href{http://arxiv.org/abs/hep-ph/0106054}{{\blue arXiv:hep-ph/0106054}}.

\bibitem{Choi:2002bb}
K.-w. Choi, K.~S. Jeong, and W.~Y. Song, ``{Operator analysis of neutrinoless
  double beta decay}'',
  \href{http://dx.doi.org/10.1103/PhysRevD.66.093007}{\blue Phys. Rev. D
  {\bfseries 66} (2002) 093007},
  \href{http://arxiv.org/abs/hep-ph/0207180}{{\blue arXiv:hep-ph/0207180}}.

\bibitem{deGouvea:2007qla}
A.~de~Gouvea and J.~Jenkins, ``{A Survey of Lepton Number Violation Via
  Effective Operators}'',
  \href{http://dx.doi.org/10.1103/PhysRevD.77.013008}{\blue Phys. Rev. D
  {\bfseries 77} (2008) 013008}, \href{http://arxiv.org/abs/0708.1344}{{\blue
  arXiv:0708.1344 [hep-ph]}}.

\bibitem{delAguila:2012nu}
F.~del Aguila, A.~Aparici, S.~Bhattacharya, A.~Santamaria, and J.~Wudka,
  ``{Effective Lagrangian approach to neutrinoless double beta decay and
  neutrino masses}'', \href{http://dx.doi.org/10.1007/JHEP06(2012)146}{\blue
  JHEP {\bfseries 06} (2012) 146}, \href{http://arxiv.org/abs/1204.5986}{{\blue
  arXiv:1204.5986 [hep-ph]}}.

\bibitem{Bonnet:2012kh}
F.~Bonnet, M.~Hirsch, T.~Ota, and W.~Winter, ``{Systematic decomposition of the
  neutrinoless double beta decay operator}'',
  \href{http://dx.doi.org/10.1007/JHEP03(2013)055}{\blue JHEP {\bfseries 03}
  (2013) 055}, \href{http://arxiv.org/abs/1212.3045}{{\blue arXiv:1212.3045
  [hep-ph]}}. [Erratum: JHEP 04, 090 (2014)].

\bibitem{Helo:2015fba}
J.~C. Helo, M.~Hirsch, T.~Ota, and F.~A. Pereira~dos Santos, ``{Double beta
  decay and neutrino mass models}'',
  \href{http://dx.doi.org/10.1007/JHEP05(2015)092}{\blue JHEP {\bfseries 05}
  (2015) 092}, \href{http://arxiv.org/abs/1502.05188}{{\blue arXiv:1502.05188
  [hep-ph]}}.

\bibitem{DeGouvea:2019wnq}
A.~De~Gouv\^ea, W.-C. Huang, J.~K\"onig, and M.~Sen, ``{Accessible
  Lepton-Number-Violating Models and Negligible Neutrino Masses}'',
  \href{http://dx.doi.org/10.1103/PhysRevD.100.075033}{\blue Phys. Rev. D
  {\bfseries 100} no.~7, (2019) 075033},
  \href{http://arxiv.org/abs/1907.02541}{{\blue arXiv:1907.02541 [hep-ph]}}.

\bibitem{Chen:2021rcv}
P.-T. Chen, G.-J. Ding, and C.-Y. Yao, ``{Decomposition of $d=9$ short-range
  $0\nu\beta\beta$ decay operators at one-loop level}'',
  \href{http://arxiv.org/abs/2110.15347}{{\blue arXiv:2110.15347 [hep-ph]}}.

\bibitem{Graf:2022lhj}
L.~Gr\'af, M.~Lindner, and O.~Scholer, ``{Unraveling the
  0\ensuremath{\nu}\ensuremath{\beta}\ensuremath{\beta} decay mechanisms}'',
  \href{http://dx.doi.org/10.1103/PhysRevD.106.035022}{\blue Phys. Rev. D
  {\bfseries 106} no.~3, (2022) 035022},
  \href{http://arxiv.org/abs/2204.10845}{{\blue arXiv:2204.10845 [hep-ph]}}.

\bibitem{Senjanovic:2011zz}
G.~Senjanovi\'c, ``{Neutrino mass: From LHC to grand unification}'',
  \href{http://dx.doi.org/10.1393/ncr/i2011-10061-8}{\blue Riv. Nuovo Cim.
  {\bfseries 34} no.~1, (2011) 1--68}.

\bibitem{Pas:1999fc}
H.~Pas, M.~Hirsch, H.~V. Klapdor-Kleingrothaus, and S.~G. Kovalenko, ``{Towards
  a superformula for neutrinoless double beta decay}'',
  \href{http://dx.doi.org/10.1016/S0370-2693(99)00330-5}{\blue Phys. Lett. B
  {\bfseries 453} (1999) 194--198}.

\bibitem{Pas:2000vn}
H.~Pas, M.~Hirsch, H.~V. Klapdor-Kleingrothaus, and S.~G. Kovalenko, ``{A
  Superformula for neutrinoless double beta decay. 2. The Short range part}'',
  \href{http://dx.doi.org/10.1016/S0370-2693(00)01359-9}{\blue Phys. Lett. B
  {\bfseries 498} (2001) 35--39},
  \href{http://arxiv.org/abs/hep-ph/0008182}{{\blue arXiv:hep-ph/0008182}}.

\bibitem{Cirigliano:2018yza}
V.~Cirigliano, W.~Dekens, J.~de~Vries, M.~L. Graesser, and E.~Mereghetti, ``{A
  neutrinoless double beta decay master formula from effective field theory}'',
  \href{http://dx.doi.org/10.1007/JHEP12(2018)097}{\blue JHEP {\bfseries 12}
  (2018) 097}, \href{http://arxiv.org/abs/1806.02780}{{\blue arXiv:1806.02780
  [hep-ph]}}.

\bibitem{Deppisch:2020ztt}
F.~F. Deppisch, L.~Graf, F.~Iachello, and J.~Kotila, ``{Analysis of light
  neutrino exchange and short-range mechanisms in $0\nu\beta\beta$ decay}'',
  \href{http://dx.doi.org/10.1103/PhysRevD.102.095016}{\blue Phys. Rev. D
  {\bfseries 102} no.~9, (2020) 095016},
  \href{http://arxiv.org/abs/2009.10119}{{\blue arXiv:2009.10119 [hep-ph]}}.

\bibitem{Cirigliano:2022oqy}
V.~Cirigliano {\em et~al.}, ``{Neutrinoless Double-Beta Decay: A Roadmap for
  Matching Theory to Experiment}'',
  \href{http://arxiv.org/abs/2203.12169}{{\blue arXiv:2203.12169 [hep-ph]}}.

\bibitem{GERDA:2020xhi}
{ GERDA}, M.~Agostini {\em et~al.}, ``{Final Results of GERDA on the Search for
  Neutrinoless Double-$\beta$ Decay}'',
  \href{http://dx.doi.org/10.1103/PhysRevLett.125.252502}{\blue Phys. Rev.
  Lett. {\bfseries 125} no.~25, (2020) 252502},
  \href{http://arxiv.org/abs/2009.06079}{{\blue arXiv:2009.06079 [nucl-ex]}}.

\bibitem{Aaboud:2018spl}
{ ATLAS}, M.~Aaboud {\em et~al.}, ``{Search for heavy Majorana or Dirac
  neutrinos and right-handed $W$ gauge bosons in final states with two charged
  leptons and two jets at $ \sqrt{s}=13 $ TeV with the ATLAS detector}'',
  \href{http://dx.doi.org/10.1007/JHEP01(2019)016}{\blue JHEP {\bfseries 01}
  (2019) 016}, \href{http://arxiv.org/abs/1809.11105}{{\blue arXiv:1809.11105
  [hep-ex]}}.

\bibitem{Schechter:1980gr}
J.~Schechter and J.~W.~F. Valle, ``{Neutrino Masses in SU(2) x U(1)
  Theories}'', \href{http://dx.doi.org/10.1103/PhysRevD.22.2227}{\blue Phys.
  Rev. D {\bfseries 22} (1980) 2227}.

\bibitem{Duerr:2011zd}
M.~Duerr, M.~Lindner, and A.~Merle, ``{On the Quantitative Impact of the
  Schechter-Valle Theorem}'',
  \href{http://dx.doi.org/10.1007/JHEP06(2011)091}{\blue JHEP {\bfseries 06}
  (2011) 091}, \href{http://arxiv.org/abs/1105.0901}{{\blue arXiv:1105.0901
  [hep-ph]}}.

\bibitem{Dorsner:2006fx}
I.~Dorsner and P.~Fileviez~Perez, ``{Upper Bound on the Mass of the Type III
  Seesaw Triplet in an SU(5) Model}'',
  \href{http://dx.doi.org/10.1088/1126-6708/2007/06/029}{\blue JHEP {\bfseries
  06} (2007) 029}, \href{http://arxiv.org/abs/hep-ph/0612216}{{\blue
  arXiv:hep-ph/0612216}}.

\bibitem{Senjanovic:2014pva}
G.~Senjanovi\'c and V.~Tello, ``{Right Handed Quark Mixing in Left-Right
  Symmetric Theory}'',
  \href{http://dx.doi.org/10.1103/PhysRevLett.114.071801}{\blue Phys. Rev.
  Lett. {\bfseries 114} no.~7, (2015) 071801},
  \href{http://arxiv.org/abs/1408.3835}{{\blue arXiv:1408.3835 [hep-ph]}}.

\bibitem{Senjanovic:2015yea}
G.~Senjanovi\'c and V.~Tello, ``{Restoration of Parity and the Right-Handed
  Analog of the CKM Matrix}'',
  \href{http://dx.doi.org/10.1103/PhysRevD.94.095023}{\blue Phys. Rev. D
  {\bfseries 94} no.~9, (2016) 095023},
  \href{http://arxiv.org/abs/1502.05704}{{\blue arXiv:1502.05704 [hep-ph]}}.

\bibitem{Mitra:2011qr}
M.~Mitra, G.~Senjanovic, and F.~Vissani, ``{Neutrinoless Double Beta Decay and
  Heavy Sterile Neutrinos}'',
  \href{http://dx.doi.org/10.1016/j.nuclphysb.2011.10.035}{\blue Nucl. Phys. B
  {\bfseries 856} (2012) 26--73}, \href{http://arxiv.org/abs/1108.0004}{{\blue
  arXiv:1108.0004 [hep-ph]}}.

\bibitem{BhupalDev:2014qbx}
P.~S. Bhupal~Dev, S.~Goswami, and M.~Mitra, ``{TeV Scale Left-Right Symmetry
  and Large Mixing Effects in Neutrinoless Double Beta Decay}'',
  \href{http://dx.doi.org/10.1103/PhysRevD.91.113004}{\blue Phys. Rev. D
  {\bfseries 91} no.~11, (2015) 113004},
  \href{http://arxiv.org/abs/1405.1399}{{\blue arXiv:1405.1399 [hep-ph]}}.

\bibitem{Li:2020flq}
G.~Li, M.~Ramsey-Musolf, and J.~C. Vasquez, ``{Left-Right Symmetry and Leading
  Contributions to Neutrinoless Double Beta Decay}'',
  \href{http://dx.doi.org/10.1103/PhysRevLett.126.151801}{\blue Phys. Rev.
  Lett. {\bfseries 126} no.~15, (2021) 151801},
  \href{http://arxiv.org/abs/2009.01257}{{\blue arXiv:2009.01257 [hep-ph]}}.

\bibitem{Li:2022cuq}
G.~Li, M.~J. Ramsey-Musolf, and J.~C. Vasquez, ``{Unraveling the left-right
  mixing using 0\ensuremath{\nu}\ensuremath{\beta}\ensuremath{\beta} decay and
  collider probes}'',
  \href{http://dx.doi.org/10.1103/PhysRevD.105.115021}{\blue Phys. Rev. D
  {\bfseries 105} no.~11, (2022) 115021},
  \href{http://arxiv.org/abs/2202.01789}{{\blue arXiv:2202.01789 [hep-ph]}}.

\bibitem{Dziewit:2021pak}
B.~Dziewit, M.~Kordiaczy\'nska, and T.~Srivastava, ``{Production of the Doubly
  Charged Higgs Boson in Association with the SM Gauge Bosons and/or Other
  $HTM$ Scalars at Hadron Colliders}'',
  \href{http://dx.doi.org/10.3390/sym13071240}{\blue Symmetry {\bfseries 13}
  no.~7, (2021) 1240}.

\bibitem{Blennow:2010th}
M.~Blennow, E.~Fernandez-Martinez, J.~Lopez-Pavon, and J.~Menendez,
  ``{Neutrinoless double beta decay in seesaw models}'',
  \href{http://dx.doi.org/10.1007/JHEP07(2010)096}{\blue JHEP {\bfseries 07}
  (2010) 096}, \href{http://arxiv.org/abs/1005.3240}{{\blue arXiv:1005.3240
  [hep-ph]}}.

\bibitem{deVries:2022nyh}
J.~de~Vries, G.~Li, M.~J. Ramsey-Musolf, and J.~C. Vasquez, ``{Light sterile
  neutrinos, left-right symmetry, and
  0\ensuremath{\nu}\ensuremath{\beta}\ensuremath{\beta} decay}'',
  \href{http://dx.doi.org/10.1007/JHEP11(2022)056}{\blue JHEP {\bfseries 11}
  (2022) 056}, \href{http://arxiv.org/abs/2209.03031}{{\blue arXiv:2209.03031
  [hep-ph]}}.

\bibitem{Dolan:2020doe}
M.~J. Dolan, T.~P. Dutka, and R.~R. Volkas, ``{Lowering the scale of Pati-Salam
  breaking through seesaw mixing}'',
  \href{http://dx.doi.org/10.1007/JHEP05(2021)199}{\blue JHEP {\bfseries 05}
  (2021) 199}, \href{http://arxiv.org/abs/2012.05976}{{\blue arXiv:2012.05976
  [hep-ph]}}.

\bibitem{delAguila:1980qag}
F.~del Aguila and L.~E. Ibanez, ``{Higgs Bosons in SO(10) and Partial
  Unification}'', \href{http://dx.doi.org/10.1016/0550-3213(81)90266-2}{\blue
  Nucl. Phys. B {\bfseries 177} (1981) 60--86}.

\bibitem{Mohapatra:1982aq}
R.~N. Mohapatra and G.~Senjanovi\'c, ``{Higgs Boson Effects in Grand Unified
  Theories}'', \href{http://dx.doi.org/10.1103/PhysRevD.27.1601}{\blue Phys.
  Rev. D {\bfseries 27} (1983) 1601}.

\bibitem{Preda:2022izo}
A.~Preda, G.~Senjanovic, and M.~Zantedeschi, ``{SO(10): A case for hadron
  colliders}'', \href{http://dx.doi.org/10.1016/j.physletb.2023.137746}{\blue
  Phys. Lett. B {\bfseries 838} (2023) 137746},
  \href{http://arxiv.org/abs/2201.02785}{{\blue arXiv:2201.02785 [hep-ph]}}.

\bibitem{Bajc:2006ia}
B.~Bajc and G.~Senjanovi\'c, ``{Seesaw at LHC}'',
  \href{http://dx.doi.org/10.1088/1126-6708/2007/08/014}{\blue JHEP {\bfseries
  08} (2007) 014}, \href{http://arxiv.org/abs/hep-ph/0612029}{{\blue
  arXiv:hep-ph/0612029}}.

\bibitem{Bajc:2007zf}
B.~Bajc, M.~Nemev\v{s}ek, and G.~Senjanovi\'c, ``{Probing seesaw at LHC}'',
  \href{http://dx.doi.org/10.1103/PhysRevD.76.055011}{\blue Phys. Rev. D
  {\bfseries 76} (2007) 055011},
  \href{http://arxiv.org/abs/hep-ph/0703080}{{\blue arXiv:hep-ph/0703080}}.

\bibitem{Mohapatra:1980qe}
R.~N. Mohapatra and R.~E. Marshak, ``{Local B-L Symmetry of Electroweak
  Interactions, Majorana Neutrinos and Neutron Oscillations}'',
  \href{http://dx.doi.org/10.1103/PhysRevLett.44.1316}{\blue Phys. Rev. Lett.
  {\bfseries 44} (1980) 1316--1319}. [Erratum: Phys.Rev.Lett. 44, 1643 (1980)].

\bibitem{Gardner:2018azu}
S.~Gardner and X.~Yan, ``{Processes that break baryon number by two units and
  the Majorana nature of the neutrino}'',
  \href{http://dx.doi.org/10.1016/j.physletb.2019.01.054}{\blue Phys. Lett. B
  {\bfseries 790} (2019) 421--426},
  \href{http://arxiv.org/abs/1808.05288}{{\blue arXiv:1808.05288 [hep-ph]}}.

\bibitem{tHooft:1975psz}
G.~'t~Hooft, ``{Magnetic Charge Quantization and Fractionally Charged
  Quarks}'', \href{http://dx.doi.org/10.1016/0550-3213(76)90031-6}{\blue Nucl.
  Phys. B {\bfseries 105} (1976) 538--547}.

\bibitem{tHooft:1974kcl}
G.~'t~Hooft, ``{Magnetic Monopoles in Unified Gauge Theories}'',
  \href{http://dx.doi.org/10.1016/0550-3213(74)90486-6}{\blue Nucl. Phys. B
  {\bfseries 79} (1974) 276--284}.

\bibitem{Polyakov:1974ek}
A.~M. Polyakov, ``{Particle Spectrum in Quantum Field Theory}'', JETP Lett.
  {\bfseries 20} (1974) 194--195.

\bibitem{kibble1976topology}
T.~W. Kibble, ``Topology of cosmic domains and strings'', Journal of Physics A:
  Mathematical and General {\bfseries 9} no.~8, (1976) 1387.

\bibitem{Zeldovich:1974uw}
Y.~B. Zeldovich, I.~Y. Kobzarev, and L.~B. Okun, ``{Cosmological Consequences
  of the Spontaneous Breakdown of Discrete Symmetry}'', Zh. Eksp. Teor. Fiz.
  {\bfseries 67} (1974) 3--11.

\bibitem{Zeldovich:1978wj}
Y.~B. Zeldovich and M.~Y. Khlopov, ``{On the Concentration of Relic Magnetic
  Monopoles in the Universe}'',
  \href{http://dx.doi.org/10.1016/0370-2693(78)90232-0}{\blue Phys. Lett. B
  {\bfseries 79} (1978) 239--241}.

\bibitem{Preskill:1979zi}
J.~Preskill, ``{Cosmological Production of Superheavy Magnetic Monopoles}'',
  \href{http://dx.doi.org/10.1103/PhysRevLett.43.1365}{\blue Phys. Rev. Lett.
  {\bfseries 43} (1979) 1365}.

\bibitem{Vilenkin:1984ib}
A.~Vilenkin, ``{Cosmic Strings and Domain Walls}'',
  \href{http://dx.doi.org/10.1016/0370-1573(85)90033-X}{\blue Phys. Rept.
  {\bfseries 121} (1985) 263--315}.

\bibitem{Rai:1992xw}
B.~Rai and G.~Senjanovi\'c, ``{Gravity and domain wall problem}'',
  \href{http://dx.doi.org/10.1103/PhysRevD.49.2729}{\blue Phys. Rev. D
  {\bfseries 49} (1994) 2729--2733},
  \href{http://arxiv.org/abs/hep-ph/9301240}{{\blue arXiv:hep-ph/9301240}}.

\bibitem{Dvali:2018txx}
G.~Dvali, C.~Gomez, and S.~Zell, ``{Discrete Symmetries Excluded by Quantum
  Breaking}'', \href{http://arxiv.org/abs/1811.03077}{{\blue arXiv:1811.03077
  [hep-th]}}.

\bibitem{Dvali:2013eja}
G.~Dvali and C.~Gomez, ``{Quantum Compositeness of Gravity: Black Holes, AdS
  and Inflation}'',
  \href{http://dx.doi.org/10.1088/1475-7516/2014/01/023}{\blue JCAP {\bfseries
  01} (2014) 023}, \href{http://arxiv.org/abs/1312.4795}{{\blue arXiv:1312.4795
  [hep-th]}}.

\bibitem{Dvali:2017eba}
G.~Dvali, C.~Gomez, and S.~Zell, ``{Quantum Break-Time of de Sitter}'',
  \href{http://dx.doi.org/10.1088/1475-7516/2017/06/028}{\blue JCAP {\bfseries
  06} (2017) 028}, \href{http://arxiv.org/abs/1701.08776}{{\blue
  arXiv:1701.08776 [hep-th]}}.

\bibitem{Dvali:2020etd}
G.~Dvali, ``{$S$-Matrix and Anomaly of de Sitter}'',
  \href{http://dx.doi.org/10.3390/sym13010003}{\blue Symmetry {\bfseries 13}
  no.~1, (2020) 3}, \href{http://arxiv.org/abs/2012.02133}{{\blue
  arXiv:2012.02133 [hep-th]}}.

\bibitem{Dvali:1997sa}
G.~R. Dvali, H.~Liu, and T.~Vachaspati, ``{Sweeping away the monopole
  problem}'', \href{http://dx.doi.org/10.1103/PhysRevLett.80.2281}{\blue Phys.
  Rev. Lett. {\bfseries 80} (1998) 2281--2284},
  \href{http://arxiv.org/abs/hep-ph/9710301}{{\blue arXiv:hep-ph/9710301}}.

\bibitem{Guth:1980zm}
A.~H. Guth, ``{The Inflationary Universe: A Possible Solution to the Horizon
  and Flatness Problems}'',
  \href{http://dx.doi.org/10.1103/PhysRevD.23.347}{\blue Phys. Rev. D
  {\bfseries 23} (1981) 347--356}.

\bibitem{Dvali:1997uq}
G.~R. Dvali, G.~Lazarides, and Q.~Shafi, ``{Mu problem and hybrid inflation in
  supersymmetric SU(2)-L x SU(2)-R x U(1)-(B-L)}'',
  \href{http://dx.doi.org/10.1016/S0370-2693(98)00145-2}{\blue Phys. Lett. B
  {\bfseries 424} (1998) 259--264},
  \href{http://arxiv.org/abs/hep-ph/9710314}{{\blue arXiv:hep-ph/9710314}}.

\bibitem{Dvali:1994ms}
G.~R. Dvali, Q.~Shafi, and R.~K. Schaefer, ``{Large scale structure and
  supersymmetric inflation without fine tuning}'',
  \href{http://dx.doi.org/10.1103/PhysRevLett.73.1886}{\blue Phys. Rev. Lett.
  {\bfseries 73} (1994) 1886--1889},
  \href{http://arxiv.org/abs/hep-ph/9406319}{{\blue arXiv:hep-ph/9406319}}.

\bibitem{Dvali:1995cc}
G.~R. Dvali and G.~Senjanovi\'c, ``{Is there a domain wall problem?}'',
  \href{http://dx.doi.org/10.1103/PhysRevLett.74.5178}{\blue Phys. Rev. Lett.
  {\bfseries 74} (1995) 5178--5181},
  \href{http://arxiv.org/abs/hep-ph/9501387}{{\blue arXiv:hep-ph/9501387}}.

\bibitem{Dvali:1996zr}
G.~R. Dvali, A.~Melfo, and G.~Senjanovi\'c, ``{Nonrestoration of spontaneously
  broken P and CP at high temperature}'',
  \href{http://dx.doi.org/10.1103/PhysRevD.54.7857}{\blue Phys. Rev. D
  {\bfseries 54} (1996) 7857--7866},
  \href{http://arxiv.org/abs/hep-ph/9601376}{{\blue arXiv:hep-ph/9601376}}.

\bibitem{Dvali:1995cj}
G.~R. Dvali, A.~Melfo, and G.~Senjanovi\'c, ``{Is There a monopole problem?}'',
  \href{http://dx.doi.org/10.1103/PhysRevLett.75.4559}{\blue Phys. Rev. Lett.
  {\bfseries 75} (1995) 4559--4562},
  \href{http://arxiv.org/abs/hep-ph/9507230}{{\blue arXiv:hep-ph/9507230}}.

\bibitem{Weinberg:1974hy}
S.~Weinberg, ``{Gauge and Global Symmetries at High Temperature}'',
  \href{http://dx.doi.org/10.1103/PhysRevD.9.3357}{\blue Phys. Rev. D
  {\bfseries 9} (1974) 3357--3378}.

\bibitem{Mohapatra:1979vr}
R.~N. Mohapatra and G.~Senjanovi\'c, ``{Broken Symmetries at High
  Temperature}'', \href{http://dx.doi.org/10.1103/PhysRevD.20.3390}{\blue Phys.
  Rev. D {\bfseries 20} (1979) 3390--3398}.

\bibitem{Mohapatra:1979bt}
R.~N. Mohapatra and G.~Senjanovi\'c, ``{High Temperature Behavior of Gauge
  Theories}'', \href{http://dx.doi.org/10.1016/0370-2693(79)90075-3}{\blue
  Phys. Lett. B {\bfseries 89} (1979) 57--60}.

\end{thebibliography}\endgroup

\end{document}